\documentclass[12pt,preprint]{aastex}

\shorttitle{The Distances of SNR W41 and HII regions}
\shortauthors{Leahy \& Tian}

\begin{document}

\title{The Distances of SNR W41 and overlapping HII regions}

\author{D.A. Leahy\altaffilmark{1}, W.W. Tian\altaffilmark{1,2}}
\altaffiltext{1}{Department of Physics \& Astronomy, University of Calgary, Calgary, Alberta T2N 1N4, Canada}
\altaffiltext{2}{National Astronomical Observatories, CAS, Beijing 100012, China; email: tww@iras.ucalgary.ca} 

\begin{abstract}
New HI images from the VLA Galactic Plane Survey show prominent absorption features 
associated with the supernovae remnant G23.3-0.3 (SNR W41).
We highlight the HI absorption spectra and the $^{13}$CO emission spectra of eight 
small regions on the face of W41, including four HII regions, three non-thermal emission
 regions and one unclassified region. The maximum velocity of absorption for W41 is 
78$\pm$2 km/s and the CO cloud at radial velocity 95$\pm$5 km/s is behind W41. Because 
an extended TeV source, a diffuse X-ray enhancement and a large molecular cloud at radial
 velocity 77$\pm$5 km/s overlap the center of W41, the kinematic distance is 3.9 to 4.5 
kpc for W41.
For the HII regions, our analysis shows that both G23.42-0.21 and G23.07+0.25 are at the 
far kinematic distances ($\sim$9.9 kpc and $\sim$ 10.6 kpc respectively) of their
 recombination-line velocities (103$\pm$0.5 km/s and 89.6$\pm$2.1 km/s respectively),
 G23.07-0.37 is at the near kinematic distance (4.4$\pm$0.3 kpc) of its recombination-line
 velocity (82.7$\pm$2.0 km/s), and G23.27-0.27 is probably at the near kinematic distance
 (4.1$\pm$0.3 kpc) of its recombination-line velocity (76.1$\pm$0.6 km/s).  
\end{abstract}


\section{Introduction}
The interaction between a supernovae remnant (SNR) and its surrounding interstellar gas 
(i.e. ionized hot gas, neutral atomic gas, and dense molecular gas) can effect the dynamics 
and physical/chemical processes in interstellar space. The interaction is considered to 
trigger several observable physical phenomena, e.g. OH (1720 MHz) shock-excited masers,
 enhanced values of the CO ($J=2-1$)/($J=1-0$) line ratio in a CO cloud, enhanced diffuse 
X-rays and very high energy $\gamma$-rays. 
These observational features have been used as direct indicators of a physical
 association between a SNR and its surrounding interstellar medium. 

Here, the emphasis is on the distance to W41 and its overlapping HII regions using 21 cm 
HI absorption spectra and CO emission spectra. These methods have been employed by Tian et al.
 (2007a) to measure the kinematic distances of the SNR G18.8+0.3 and its associated CO clouds.
 Gordon et al. (1976) first estimated distance of 4.6 - 7 kpc for W41 by observations of HI
 emission and absorption in the W41 region. Tian et al (2007b) revised the distance using an 
incorrect association with a molecular cloud at about 63 km/s. Generally, HI observations may 
decide distances (or distance range) of the associations and distinguish if the associations 
are physical or only morphological. However, there exists the so-called kinematic distance
 ambiguity in the inner Galaxy, i.e. each radial velocity along a given line-of-sight 
corresponds to two distances equally spaced on either side of the tangent point. CO 
observations have helped to reduce the ambiguity to some degree, since a CO cloud behind a 
SNR will not produce HI absorption feature in the SNR's absorption spectrum. 
I.e, if a CO cloud's radial velocity is more than the maximum HI absorption line velocity in 
the direction of a SNR, the SNR should be located at the near distance of the maximum HI 
absorption line velocity. Especially, if a HII region is overlapping with a SNR, the HI and 
CO observations together may solve the distance ambiguity completely: the comparison of the 
maximum HI absorption line velocity with the recombination-line velocity of the HII region may 
decide whether the HII region is at the far or near distance of the recombination-line velocity 
(e.g. Kolpak et al. 2003). In this paper, we use these methods to directly measure distances of
the SNR W41 and its overlapping HII regions, therefore verify a W41/CO cloud interaction 
which has been used to explain the origin of the TeV source and diffuse X-rays in the center of 
W41 (Tian et al. 2007b). We identify a different CO cloud, at 77 km/s, interacting with W41 
and revise the distance to W41.

\section{Radio Observations}
The HI-line data sets come from the VLA Galactic Plane Survey (VGPS, Stil et al. 2006). The
 data sets are mainly based on observations from VLA of the National Radio Astronomy 
Observatory (NRAO). The synthesized beam for the HI line images is 1$^{\prime}$, the radial 
velocity resolution is 1.56 km$/$s, and the rms noise is 2 K per channel.  The short-spacing 
information for the H I spectral line images is from additional observations with the 100 m 
Green Bank Telescope of the NRAO. The CO-line ($J=1-0$) data set is from the Galactic ring 
survey (Jackson et al. 2006) by employing the Five College Radio Astronomy Observatory 14 m 
telescope. The CO-line data in the paper have velocity coverage of -5 to 135 km/s, an angular 
sampling of 22$^{\prime\prime}$, radial velocity resolution of 0.21 km/s, and rms noise of 
$\sim$0.13 K. 
 
\section{Results}

\subsection{HI Emission and Absorption Spectra}
We have searched the VGPS radial velocity range from -110 to 170 km/s for features in the HI 
which might be related to the morphology of W41. There are HI features coincident with the SNR 
in three velocity ranges: 0 to 10 km/s, 50 to 63 km/s, and 69 to 79 km/s (see Fig. 1). These 
maps have superimposed contours (at 30, 45, 62, 148 K) of 1420 MHz continuum emission chosen to 
show the SNR. Clear absorption features are detected in almost the whole radial velocity range  
0 to 106 km/s, which are associated with the brightest HII region G23.42-0.21 overlapping W41.
 These HI features are strongly correlated with the continuum intensity, indicating that they 
are caused by HI absorption between the continuum emission source and the earth. 

In order to study the distances of W41 and its overlapping HII regions, we obtained their HI
 absorption spectra. The methods to construct HI absorption spectra are discussed in general by 
Dickey and Lockman (1990). For extended sources like W41, the methods have been extended by
 Tian et al. (2007a), i.e. 
  $\Delta T$ = $T^{HI}_{off}$-$T^{HI}_{on}$ = ($T^{c}_{s}$-$T^{c}_{bg}$)(1-$e^{-\tau}$).
$T^{HI}_{on}$ and $T^{HI}_{off}$ are the average brightness temperatures of many spectra from 
a selected area on a strong continuum emission region of the source and of an adjacent 
background region. Here we use a box for the source region and an annular box excluding the 
source box.  $T^{c}_{s}$ and $T^{c}_{bg}$ are the average continuum brightness temperatures 
for the same regions respectively.  $\tau$ is the optical depth from the continuum source to 
the observer along the line-of-sight. 
 Eight small source regions, marked in the upper left plot of Fig. 2, have been selected from 
the W41 area, including four thermal HII regions which have been identified before, three 
non-thermal emission regions originating from W41, and one unclassified region which is 
probably a HII region due to its strong continuum emission at 5 GHz (named as G23.24-0.11, 
see Becker et al. 1994). The seven other plots of Fig. 2 show the HI emission and absorption 
spectra of these regions. 
The absorption features mostly line up with valleys in the emission spectra since the 
background spectra also cover part of continuum emission regions. 
Table 1 summarizes the parameters used to obtain the spectra. To estimate uncertainties of 
the absorption spectra, we calculate the standard derivation $\sigma$ of e$^{-\tau}$ for 
velocities $>$ 120 km/s and $<$ -40 km/s. The values of 3$\sigma$ are listed in Table 1. 
This gives some indicator of what size of errors to expect. However, since the features
are mainly not statistical but rather due to real clouds, that happen to be either in the
source box (or the background box) but not in the background box (or the source box), this is 
only a rough indicator of what is spurious.
The maximum radial velocity of absorption in the direction of each region and the available
 recombination-line velocities of the four HII regions are summarized in Table 2. 

\subsection{CO Emission Spectra}
Fig. 3 shows $^{13}$CO-line emission spectra of the eight regions. The $^{13}$CO emission 
spectra over the full area of W41 in velocity range of -5 to 135 km/s show five high
 brightness-temperature CO molecular clouds, at radial velocities 53$\pm$3, 63$\pm$4, 77$\pm$5,
 95$\pm$5 and 101$\pm$5 km/s respectively, and two low brightness-temperature CO molecular 
clouds, at 4$\pm$1 km/s and 20$\pm$3 km/s respectively. The third line of Table 2 gives the 
radial velocity of the CO cloud which corresponds to the maximum HI absorption for each of the 
eight regions. The fourth line gives the radial velocity of the CO cloud behind the CO cloud 
corresponding to the maximum HI absorption: this velocity gives an  upper limit to the distance 
since the continuum emission must be located in front of this CO cloud.  

\section{Discussion}  
\subsection{The interaction between W41 and its surrounding CO clouds}
 Albert et al. (2006) studied the $^{12}$CO data and found a giant molecular cloud (GMC) with 
a $^{12}$CO peak emission from 70 to 85 km/s. Our $^{13}$CO emission spectra show that a GMC 
at radial velocity of 77$\pm$5 km/s is detected in all eight emission regions including the 
center of W41. Recent observations detected a diffuse X-ray enhancement likely associated 
with an extended TeV source in the center of W41 (Tian et al. 2007b). The 77$\pm$5 km/s CO 
cloud is likely the one 
responsible for the TeV emission by TeV protons accelerated by the SNR colliding with protons 
in the cloud to produce pions, which decay to $\gamma$-rays (Yamazaki et al. 2006). Therefore, 
W41 is associated with this GMC physically, i.e., W41 is adjacent to or within the cloud at 
77$\pm$5 km/s, and their interaction is responsible for the detected X-rays, likely due to 
synchrotron from secondary electrons, and $\gamma$-rays, due to pion decay. In Tian et al. 
(2007b), the distance to W41 and association with a CO cloud at $\sim$63 km/s were incorrect 
due to association of W41 with a HI absorption feature at this velocity. 
We still conclude that W41 interacts with a CO cloud, but the cloud is at 77$\pm$5 km/s.

\subsection{Distance of W41 and the overlapping HII regions}
The HI emission spectra in Fig. 2 indicate a tangent point velocity of $V_{t}$= 112$\pm$2 km/s 
in the direction of W41. A pure circular velocity model can not be used to obtain V$_{R}$ due 
to non-circular motions (e.g. Brand \& Blitz 1993). $V_{t}$=112$\pm$2 km/s is consistent with 
the value from fig. 8 of  Weiner $\&$ Sellwood (1999) who include non-circular motions.
W41's distance is limited by the maximum radial velocity for absorption of 78$\pm$2 km/s 
from the absorption spectra of W41 (regions C1, 2 and 7). Also, the CO cloud (95$\pm$5 km/s) is 
behind W41, because it shows no respective HI absorption feature due to W41 
(regions C1, 2 and 7). 
There are two noisy absorption features at 95 km/s and -30 km/s respectively in the region 
2 spectrum. 
Also the continuum emission of region 2 is partly due to a weak HII region G23.42-0.39. 
So we think the two absorption features are probably either not real due to limited accuracy 
of the background HI spectrum or else related with the adjacent G23.42-0.39 (see below 
for detail), which is so faint in Fig. 1 that it is difficult for us to obtain a 
reliable absorption spectrum here.  
Since W41 is closer than the tangent point and V$_{R}$ (78$\pm$2 km/s) is significantly 
less than $V_{t}$, W41 is close enough to use the Galaxy velocity field of Brand $\&$ 
Blitz (1993). Recent measurements of the parameters for the Galactic rotation curve model 
give R$_{0}$=7.6$\pm$0.3 kpc (Eisenhauer et al. 2005) and V$_{0}$=214$\pm$7 km/s 
(Feast \& Whitelock 1997, Reid \& Brunthaler 2004). 
Figs. 6a and 6b of Brand $\&$ Blitz (1993) show that the mean rotation curve is nearly 
flat for these values, so we use a flat rotation curve. For pure circular rotation, the 
distance to W41 is d $\sim$ 4.4 kpc (R $\sim$ 4.0 kpc). Fig. 10 of Brand $\&$ Blitz (1993)
 shows a systematic velocity deviation of $\sim$+5 km/s from circular rotation 
in the direction of W41 (also consistent with Weiner $\&$ Sellwood 1999) for d $\sim$ 4.4  
kpc. Using a systematic deviation of $\sim$ +5 km/s and a random velocity of $\pm$5 km/s, 
the radial velocity of W41 due to circular rotation is 66 to 80 km/s. This yields a 
distance to W41 of d= 3.9 to 4.5 kpc (R= 4.3 to 3.9 kpc). 

The recombination-line velocities of the four HII regions have been obtained previously
(Lockman 1989). By comparison of the HI absorption spectrum with the CO emission spectrum, 
we find all major CO clouds have respective HI absorption in the direction of the bright 
HII region G23.42-0.21 (region 3). 
Kolpak et al (2003) showed an HI absorption spectrum of G23.42-0.21 and suggested the 
distance to G23.42-0.21 (with low confidence) is at the near side of the 
recombination line velocity since the HII recombination line velocity, maximum absorption 
velocity and tangent point velocity are nearly the same. We obtained a similar absorption 
spectrum with the additional information of
the CO emission feature velocity, but clear absorption features at 106 km/s in the 
HI map are associated with G23.42-0.21. We argue the maximum absorption velocity exceeds 
the recombination line velocity, so G23.42-0.21 is at the far kinematic distance of the 
recombination-line velocity (103$\pm$0.5 km/s), i.e. 9.9$\pm$0.3 kpc (here we use the 
systematic velocity field given by Weiner $\&$ Sellwood 1999). 
Similarly, the maximum absorption velocity of HII region G23.07-0.25 (region 5) is more than 
its respective recombination-line velocity. This shows that this HII region is also 
located at the far kinematic distance of its recombination-line velocity, i.e. 
10.6$\pm$0.3 kpc for 89.6$\pm$2.1 km/s.  

The small emission region G23.42-0.39, adjacent to the region 2, has been listed in a new 
compact HII region catalog (Giveon et al. 2005) based on its spectral index between 
1.4 and 5 GHz, but no recombination-line is detected. Because of its faint continuum emission 
at 1.4 GHz (Fig. 1), we can not obtain a reliable absorption spectrum, nor a distance. 
HII region G23.07-0.37 (region 6) is at the near kinematic distance (4.4$\pm$0.3 kpc) of its 
recombination-line velocity 82.7$\pm$2.0 km/s, since there is no absorption seen beyond this 
velocity. For HII region G23.27-0.27 (region C2), a weak absorption feature at 95$\pm$2 km/s 
seems to be related with a CO emission at 95$\pm$5 km/s. However, another CO emission feature 
at 103$\pm$2 km/s clearly doesn't produce respective HI absorption, so we think the weak 
absorption feature at 95$\pm$2 km/s is not real. In this case,  G23.27-0.27 is probably at the 
near distance (4.1$\sim$ 0.3 kpc) of its recombination-line velocity 76.1$\pm$0.6 km/s. 
Although the unclassified source G23.24-0.11 (region 4) lies on the bright arc in the northern 
border of W41, its maximum absorption velocity is at 102 km/s, which means the source is 
not part of the SNR. 

\section{Conclusion}
The HI absorption spectra and the CO emission spectra of four
 HII regions, three non-thermal emission regions and one unclassified region covering 
W41 have been constructed. They show that
the maximum radial velocity of absorption for W41 is 78$\pm$2 km/s and the CO cloud at 
 95$\pm$5 km/s is behind W41. Further considering the evidence that an extended 
TeV source, a diffuse X-ray enhancement and a GMC at 77$\pm$5 km/s are also 
located at the center of W41, we think that W41 is associated with the GMC physically. 
Thus we obtain a kinematic distance of 3.9 to 4.5 kpc for W41. In addition, by combining HI 
absorption features with known recombination-line velocities of four HII regions overlapping 
W41, we give the distances of the four HII regions.

\begin{acknowledgements}
DAL and WWT acknowledge support from the Natural Sciences and Engineering Research Council of Canada.
WWT thanks support from the Natural Science Foundation of China.  
This publication makes use of molecular line data from the Boston University-FCRAO Galactic Ring Survey (GRS). The NRAO is a facility of the National Science Foundation operated under cooperative agreement by Associated Universities, Inc. 
\end{acknowledgements}

\clearpage

\begin{table}
\begin{center}
\caption{Parameters of Eight HI Spectra}
\setlength{\tabcolsep}{1mm}
\begin{tabular}{ccccccccc}
\hline
 Region:  &  C1 & C2 & 2 & 3 & 4 & 5 & 6& 7 \\
\hline
 Coordinate l:  & 23.25& 23.27& 23.46&23.42 & 23.24& 23.07& 23.07& 23.12\\
  Coordinate b:  & -0.24&-0.27 &-0.35 &-0.21& -0.11& -0.25&-0.37&-0.50\\
 No. source spectra     & 49  & 126  & 345  & 322  & 210 & 360 & 437 & 232\\
 No. background spectra & 91  & 216  & 728  & 935  & 726 & 663 & 719 & 667\\
 $T^{c}_{s}$(K)  & 62.4 & 57.4& 34.6 & 83.8 & 49.3 & 57.9& 53.7& 36.7 \\
 $T^{c}_{bg}$(K) & 41.9 & 41.1& 27.1 & 33.9 & 28.9 & 42.7& 39.1& 25.9 \\
 Standard derivation(3$\sigma$)& 0.17&0.18&0.22&0.08&0.13&0.14&0.11&0.12\\
\hline
\hline
\end{tabular}
\end{center}
\end{table}

\begin{table}
\begin{center}
\caption{Summary of HI absorption and CO emission features}
\setlength{\tabcolsep}{1mm}
\begin{tabular}{ccccccccc}
\hline
Region Number: &  C1 & C2 & 2 & 3 & 4 & 5  & 6 & 7 \\
km/s & SNR &HII&SNR&HII&?&HII&HII&SNR \\
\hline
\hline
Maximum absorption velocity:  & 76$\pm$2 & 95$\pm$2 & 78$\pm$2 & 104$\pm$2 & 102$\pm$1 & 100$\pm$2 & 75$\pm$3& 78$\pm$2 \\
HII recombination-line velocity: &  & 76.1$\pm$0.6 & none & 103$\pm$0.5& & 89.6$\pm$2.1 & 82.7$\pm$2.9& \\
Nearby CO emission feature:  & 77$\pm$5 & 95$\pm$5 & 78$\pm$2 & 102$\pm$5 & 95$\pm$5& 102$\pm$2 & 76$\pm$3& 77$\pm$5\\
Far CO emission feature:  & 95$\pm$5 &  & 102$\pm$5 &  & 102$\pm$2&  & 84$\pm$3& \\
\hline
\hline
\end{tabular}
\end{center}
\end{table}

\clearpage

\begin{figure}
\includegraphics[angle=-90,width=.5\textwidth]{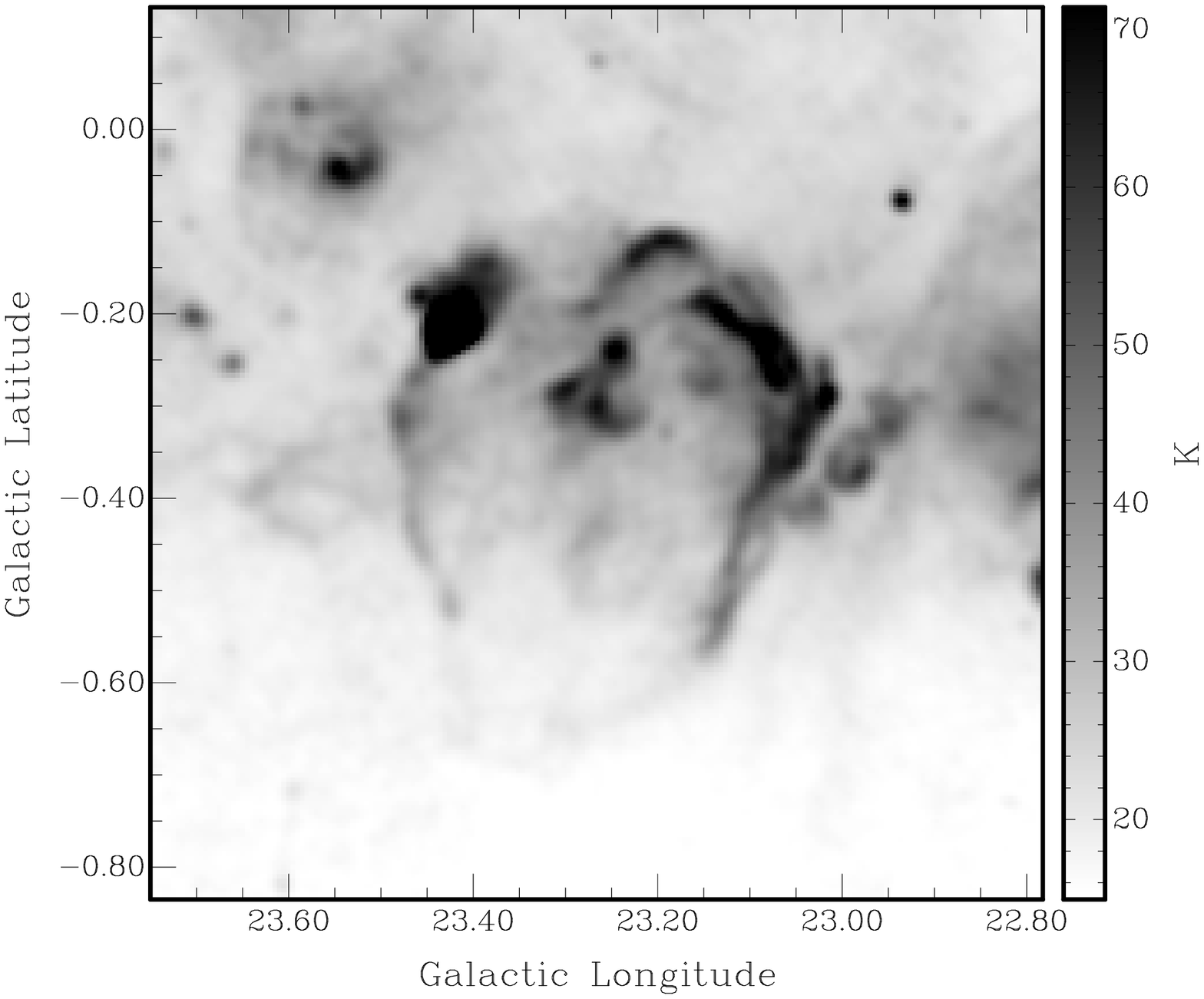}\includegraphics[angle=-90,width=.5\textwidth]{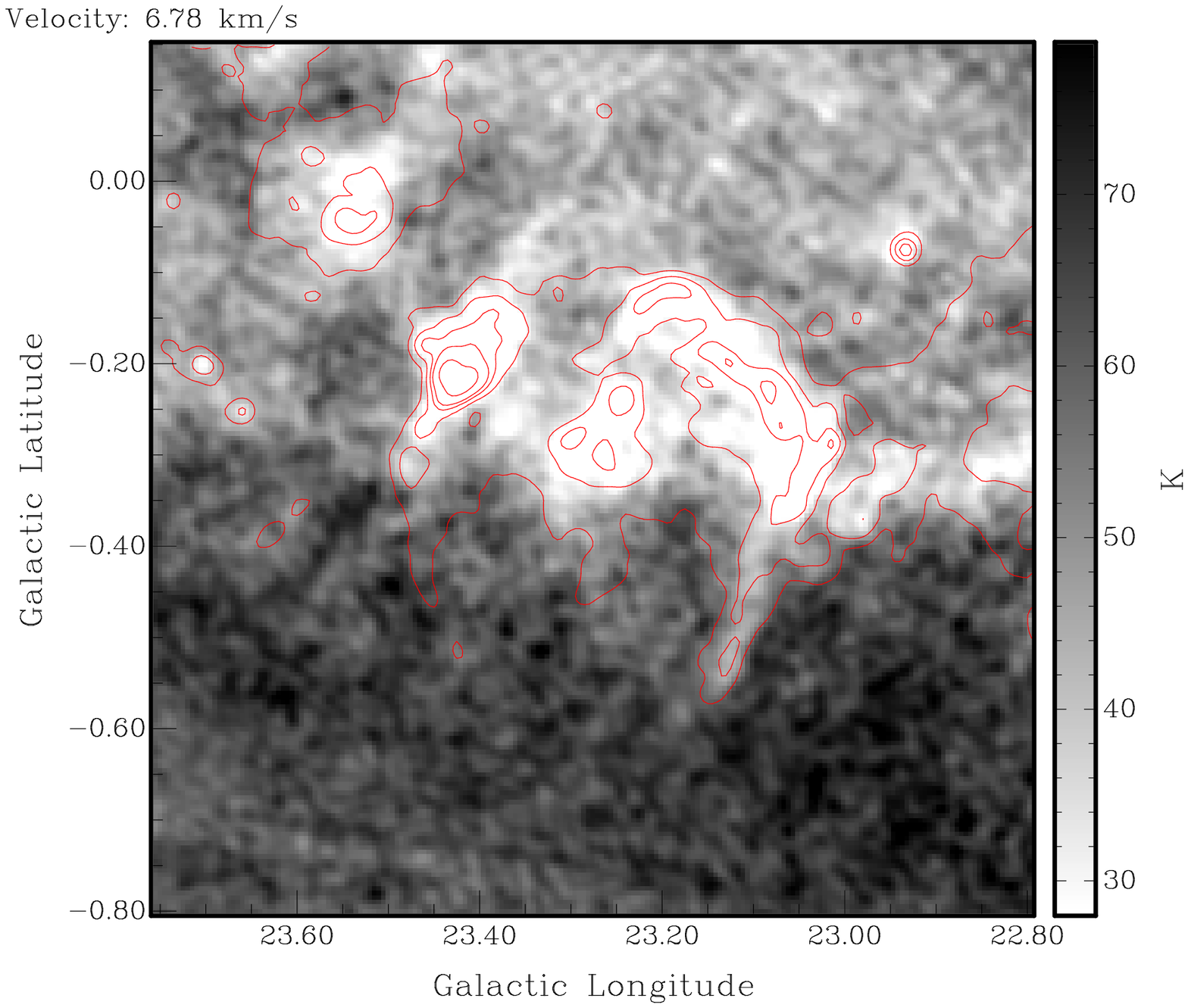}\\
\includegraphics[angle=-90,width=.5\textwidth]{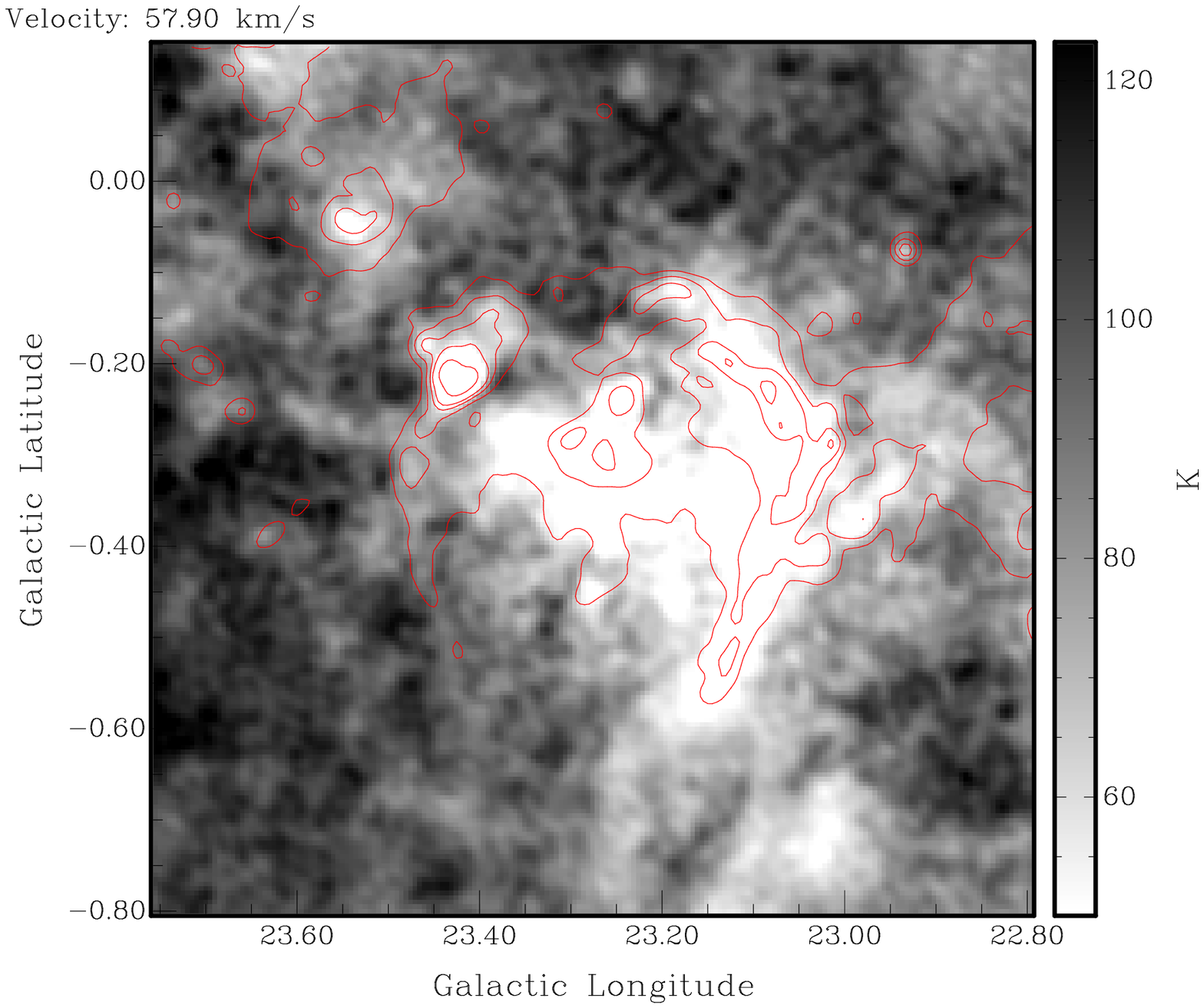}\includegraphics[angle=-90,width=.5\textwidth]{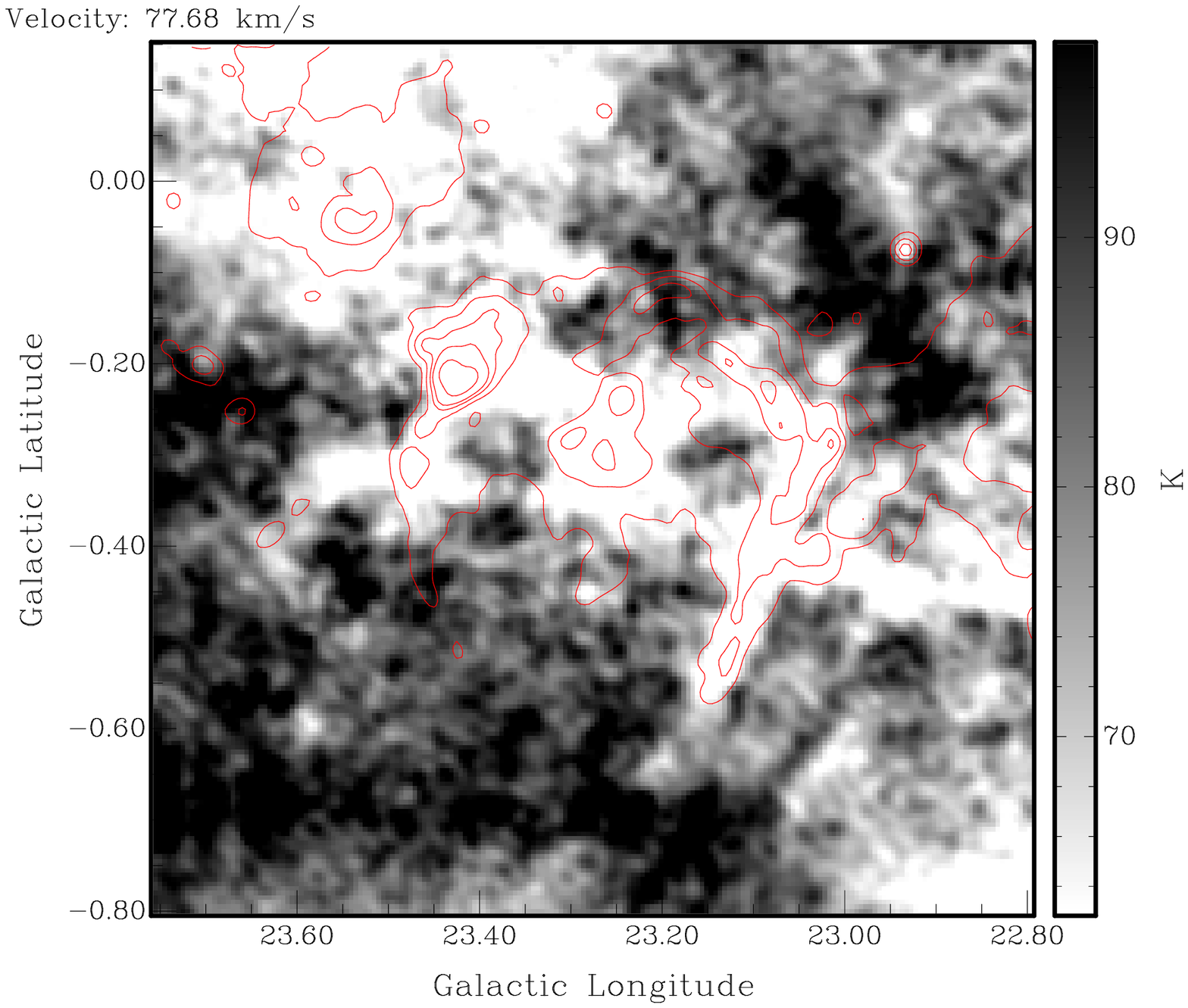}
\caption{The 1420 MHz continuum image (upper left) of W41 and the images of HI emission at three channels: 7, 58 and 78 km/s respectively. The HI maps have superimposed contours (30, 45, 62, 148 K, green) of 1420 MHz continuum emission to show the SNR and HII regions.}
\end{figure}

\begin{figure}
\vspace{185mm}
\begin{picture}(80,80)
\put(20,460){\includegraphics{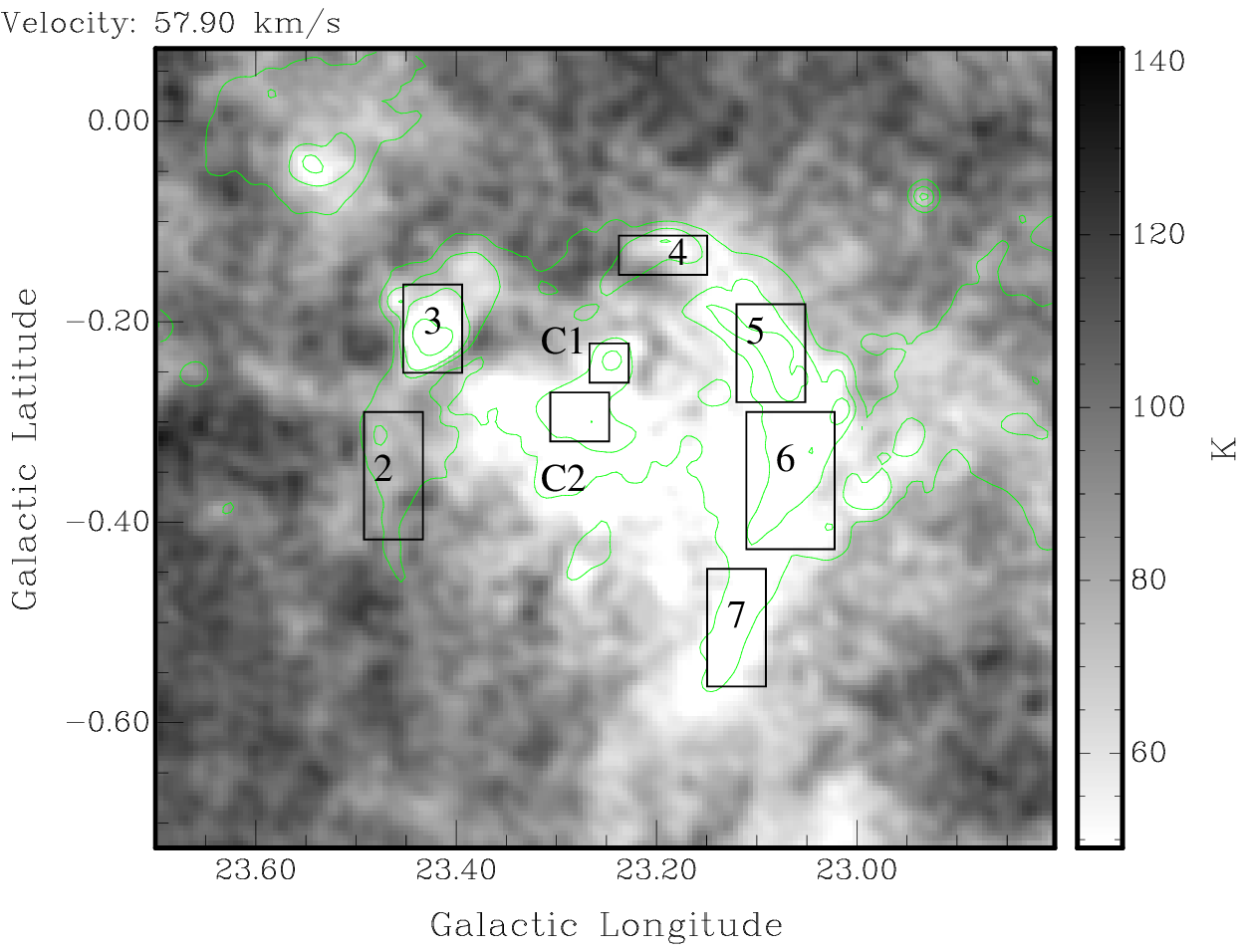}}
\put(260,455){\includegraphics{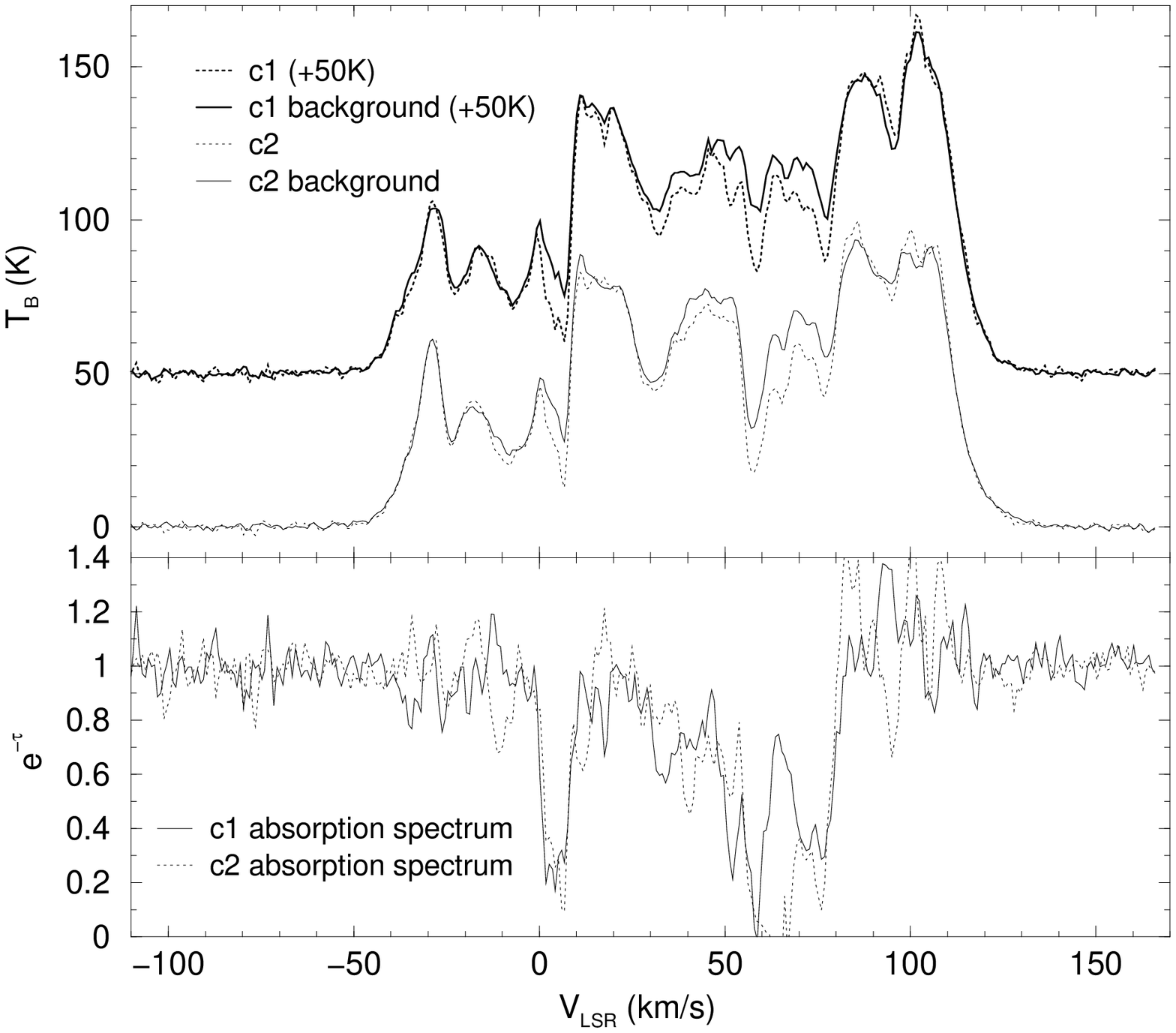}}
\put(0,305){\includegraphics{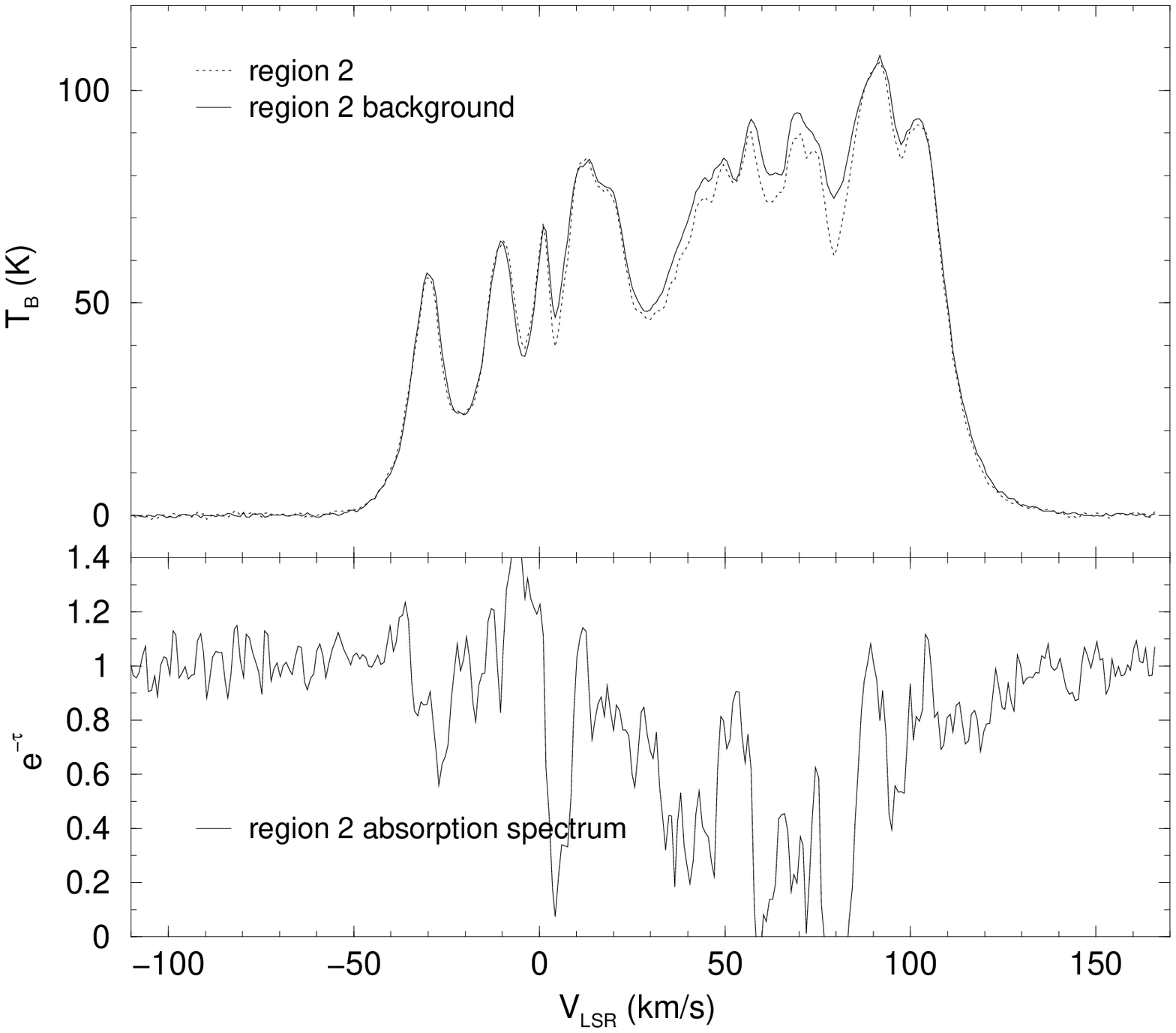}}
\put(260,305){\includegraphics{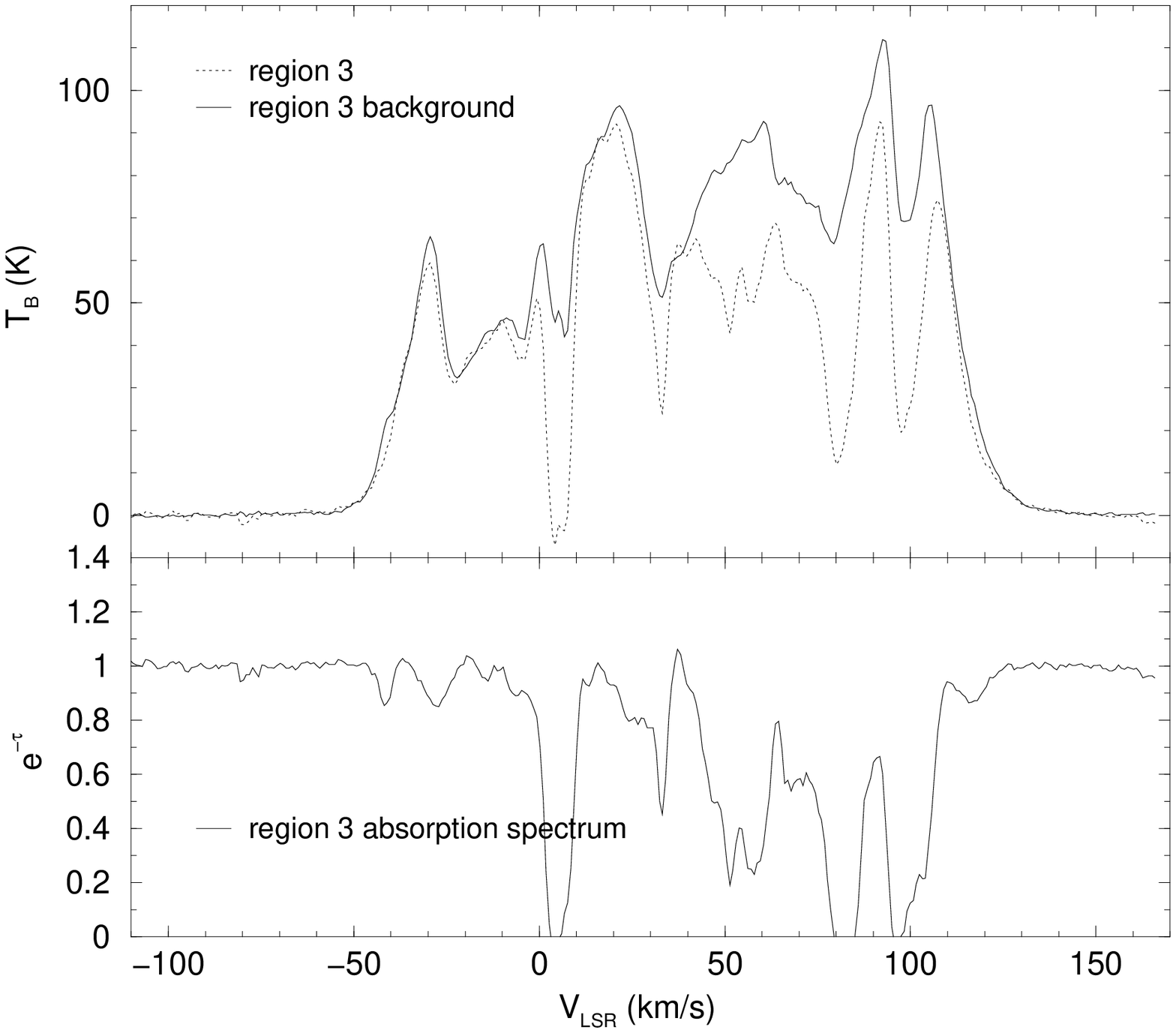}}
\put(0,150){\includegraphics{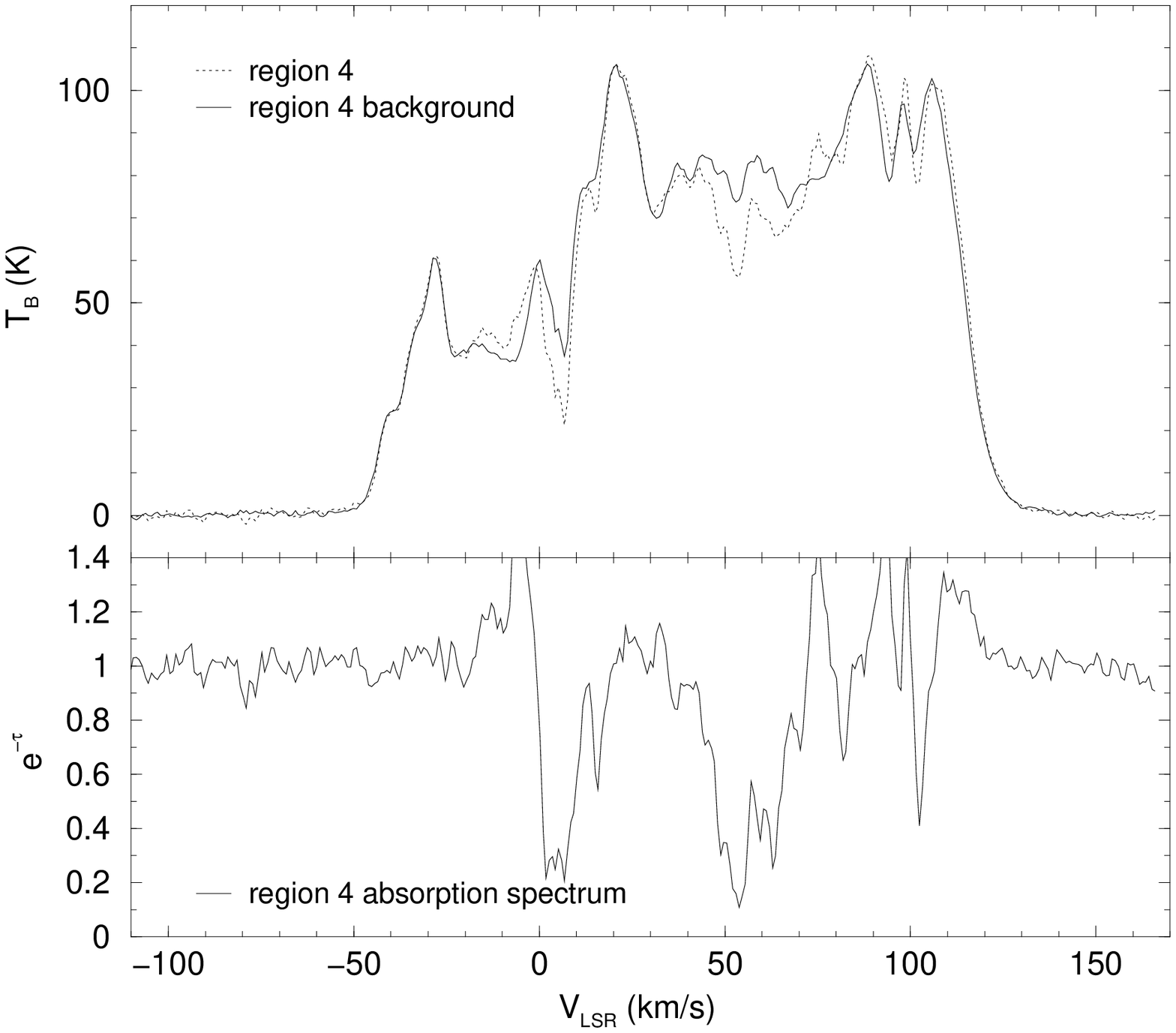}}
\put(260,150){\includegraphics{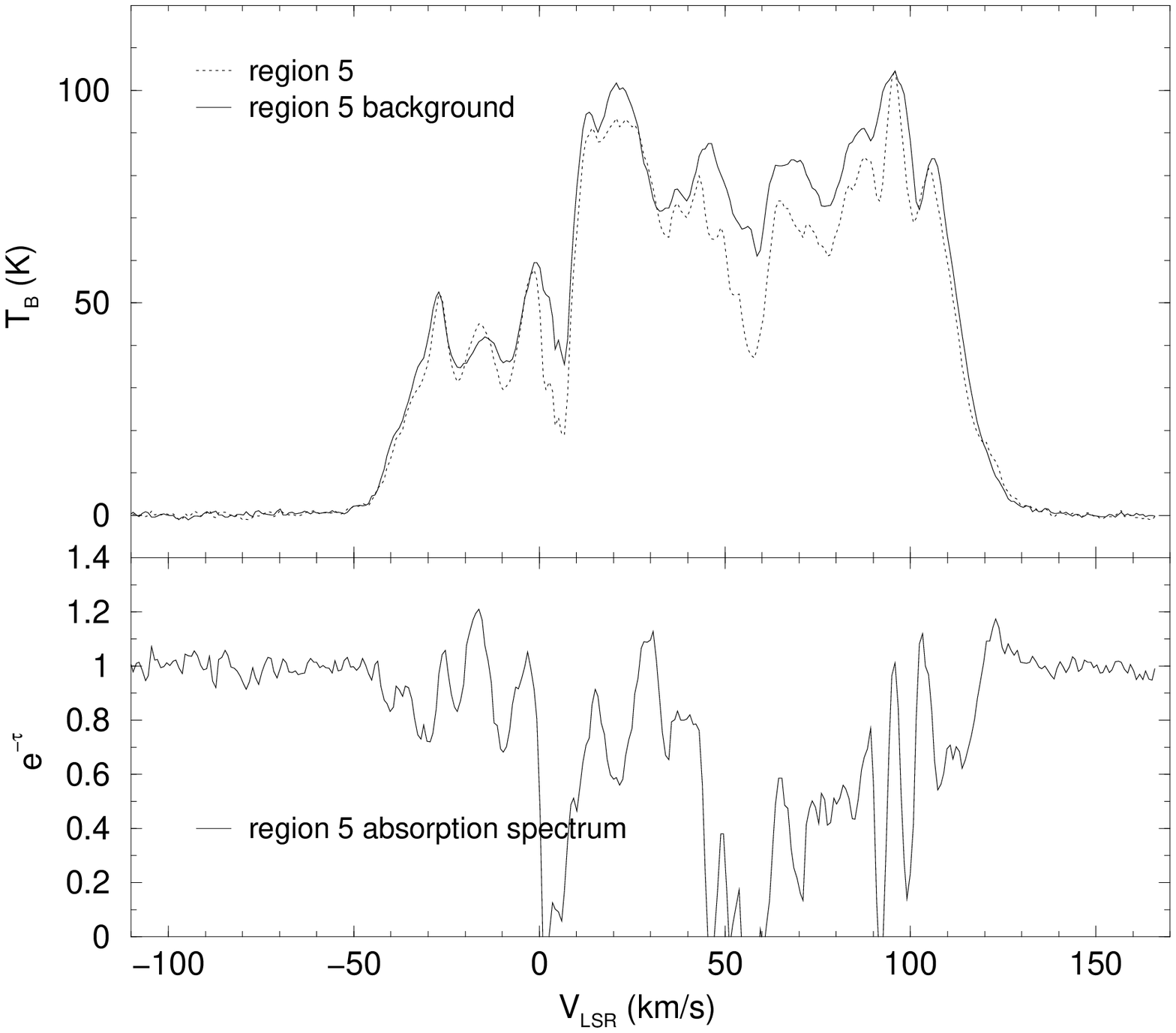}}
\put(0,-5){\includegraphics{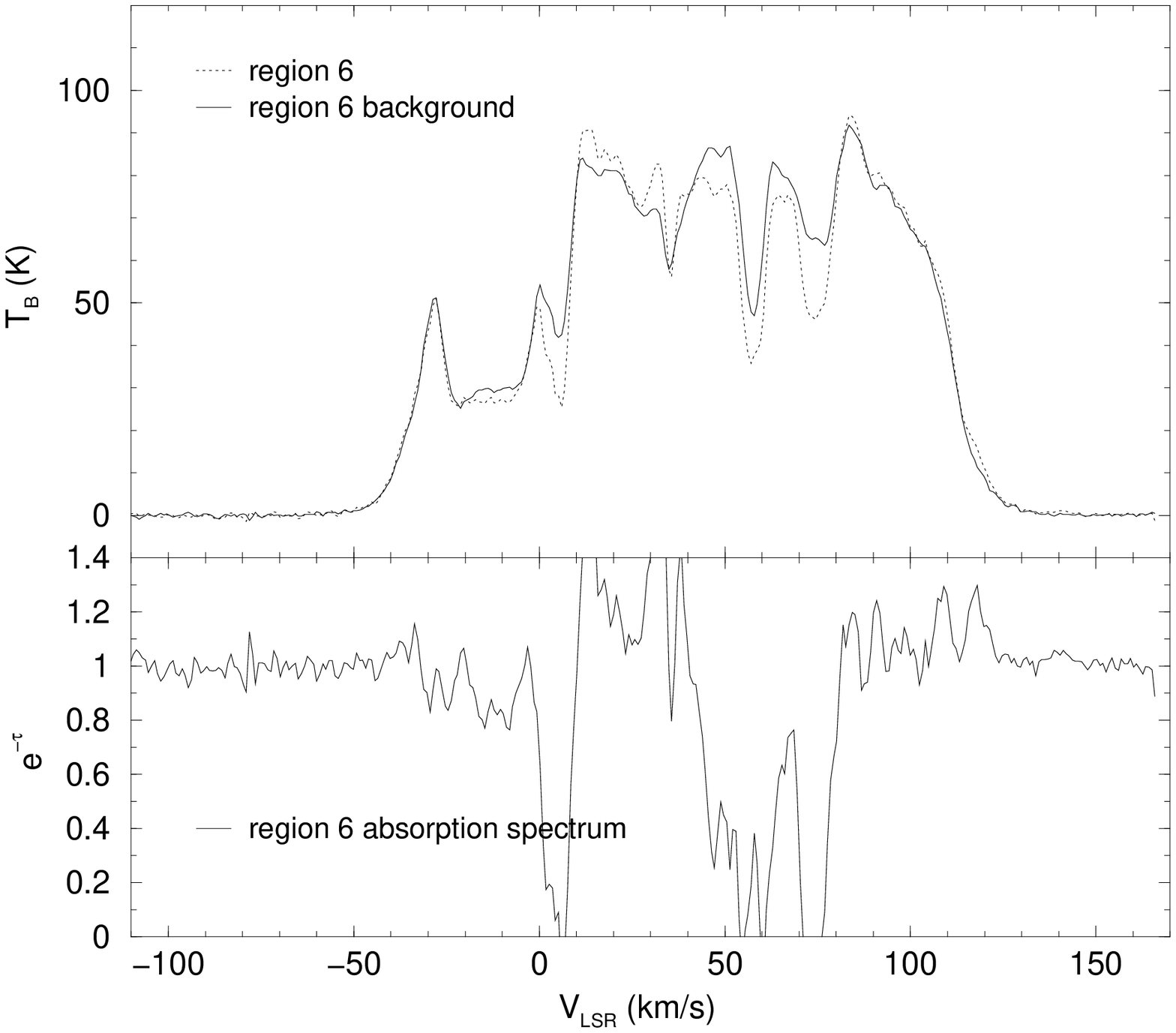}}
\put(260,-5){\includegraphics{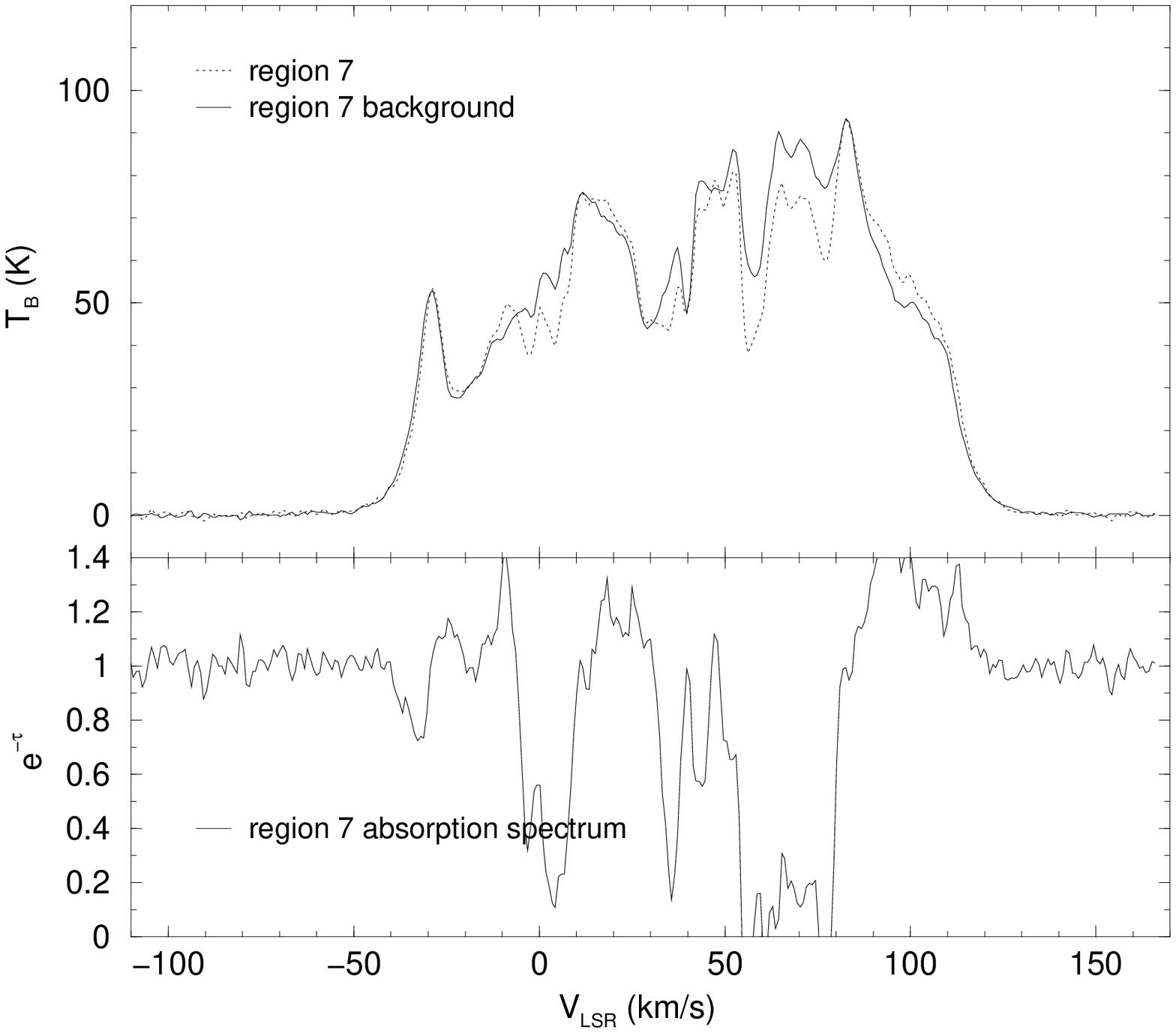}}
\end{picture}
\caption{ HI image of W41 from a single channel at 57.9 km/s (upper left). The seven other panels show the HI spectra, extracted from boxes C1, C2, 2, 3, 4, 5, 6 and 7 shown in the first panel.}
\end{figure}

\begin{figure}
\vspace{190mm}
\begin{picture}(80,80)
\put(0,650){\includegraphics{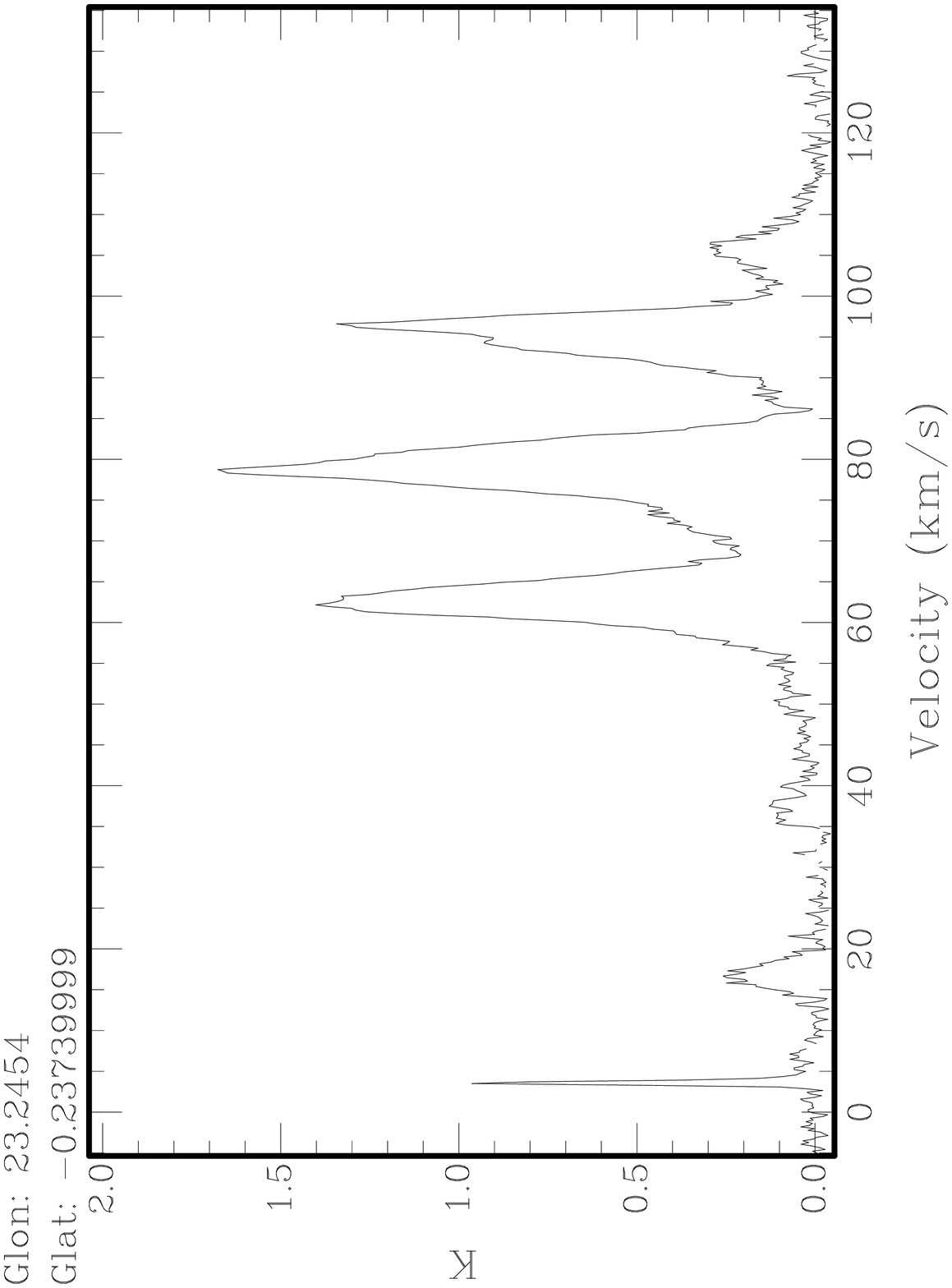}}
\put(260,650){\includegraphics{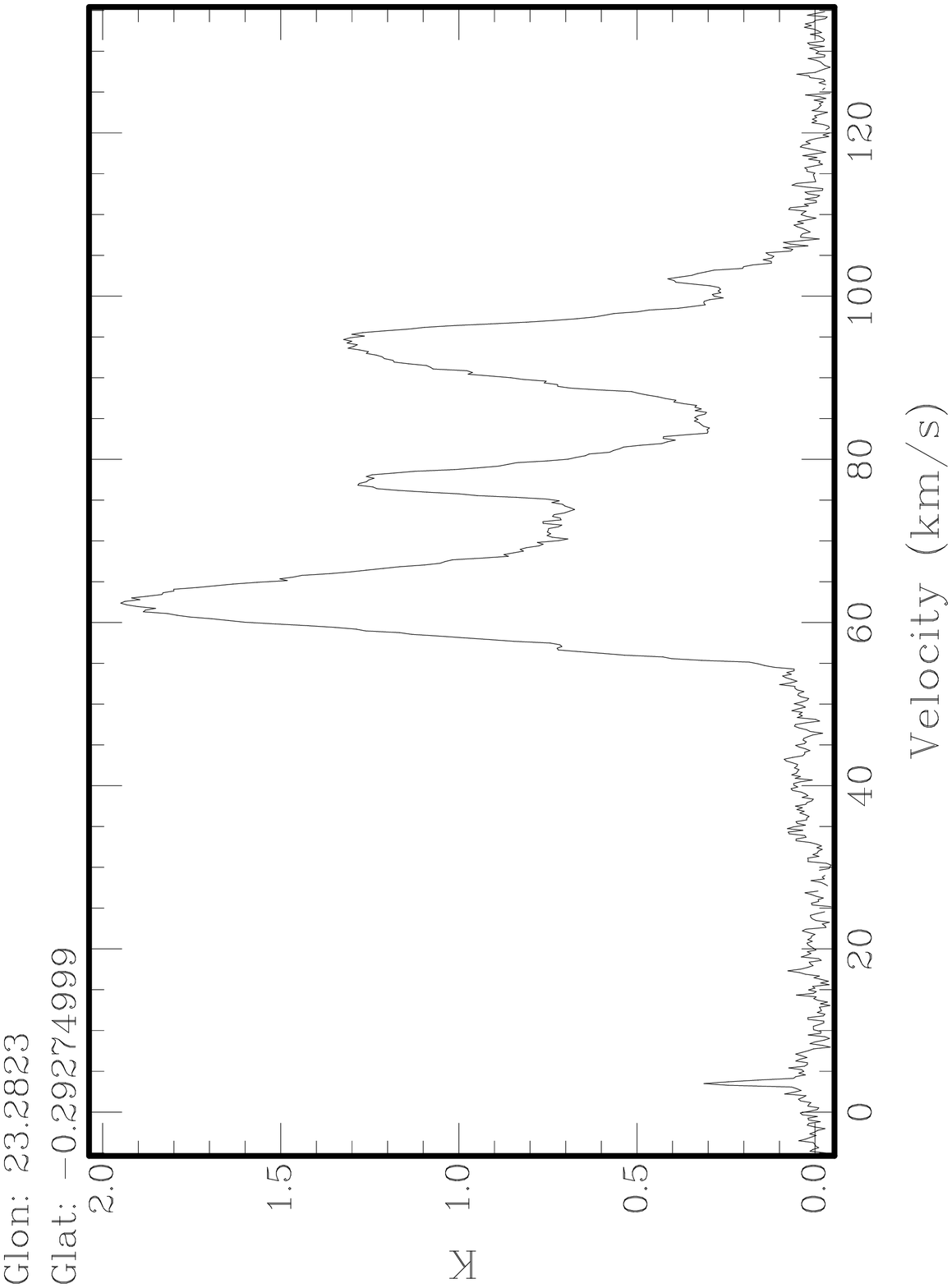}}
\put(0,485){\includegraphics{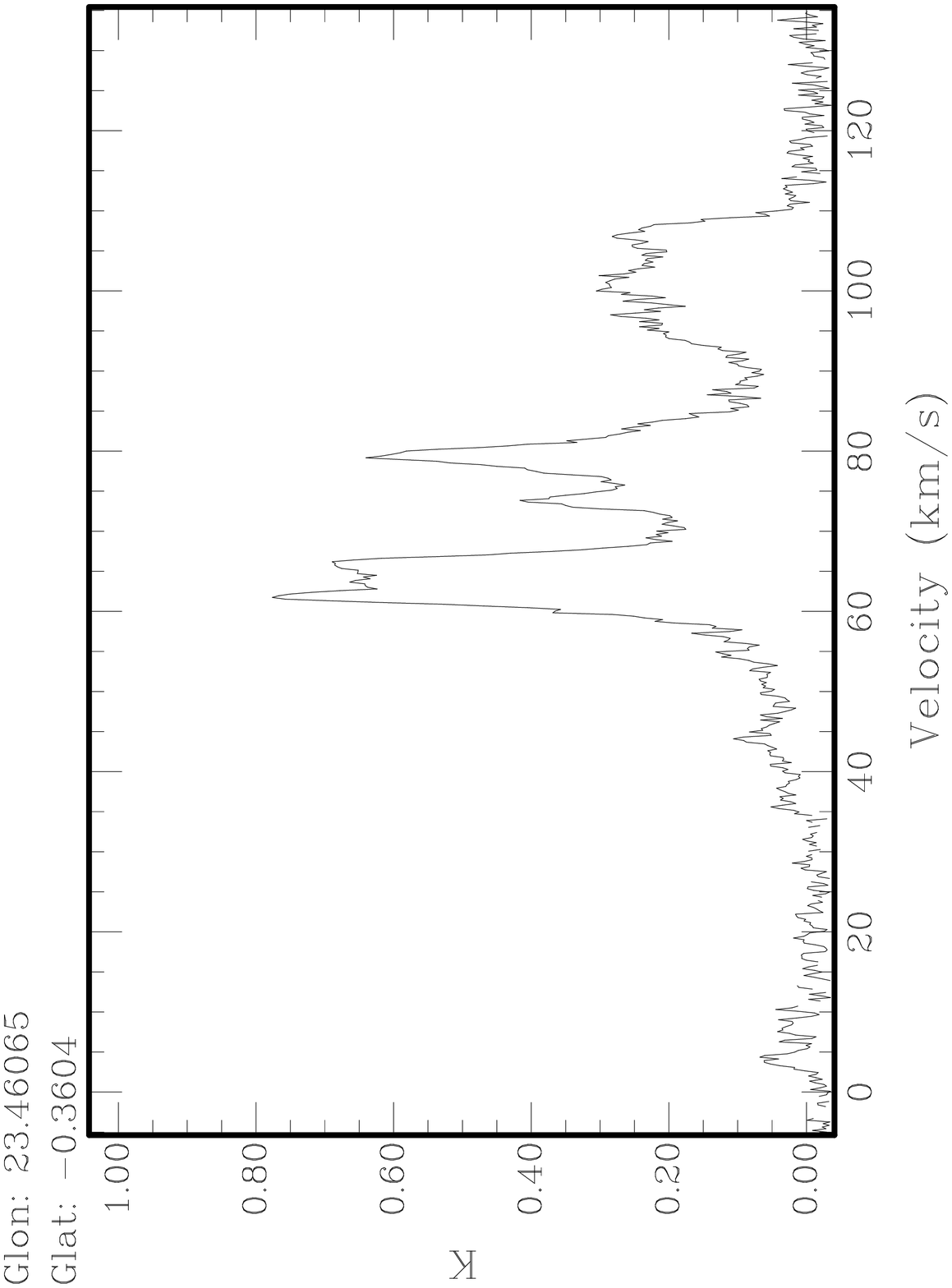}}
\put(260,485){\includegraphics{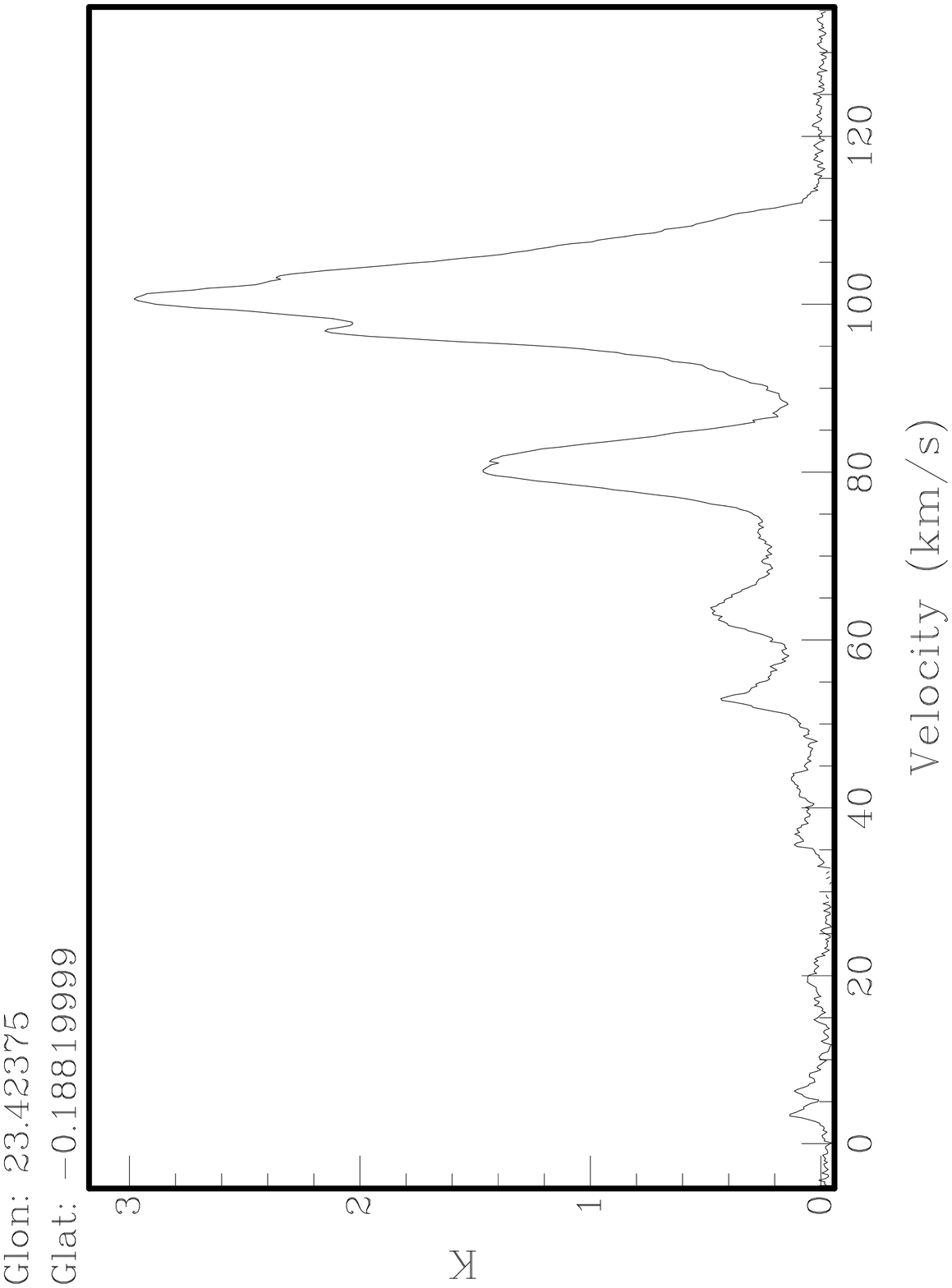}}
\put(0,320){\includegraphics{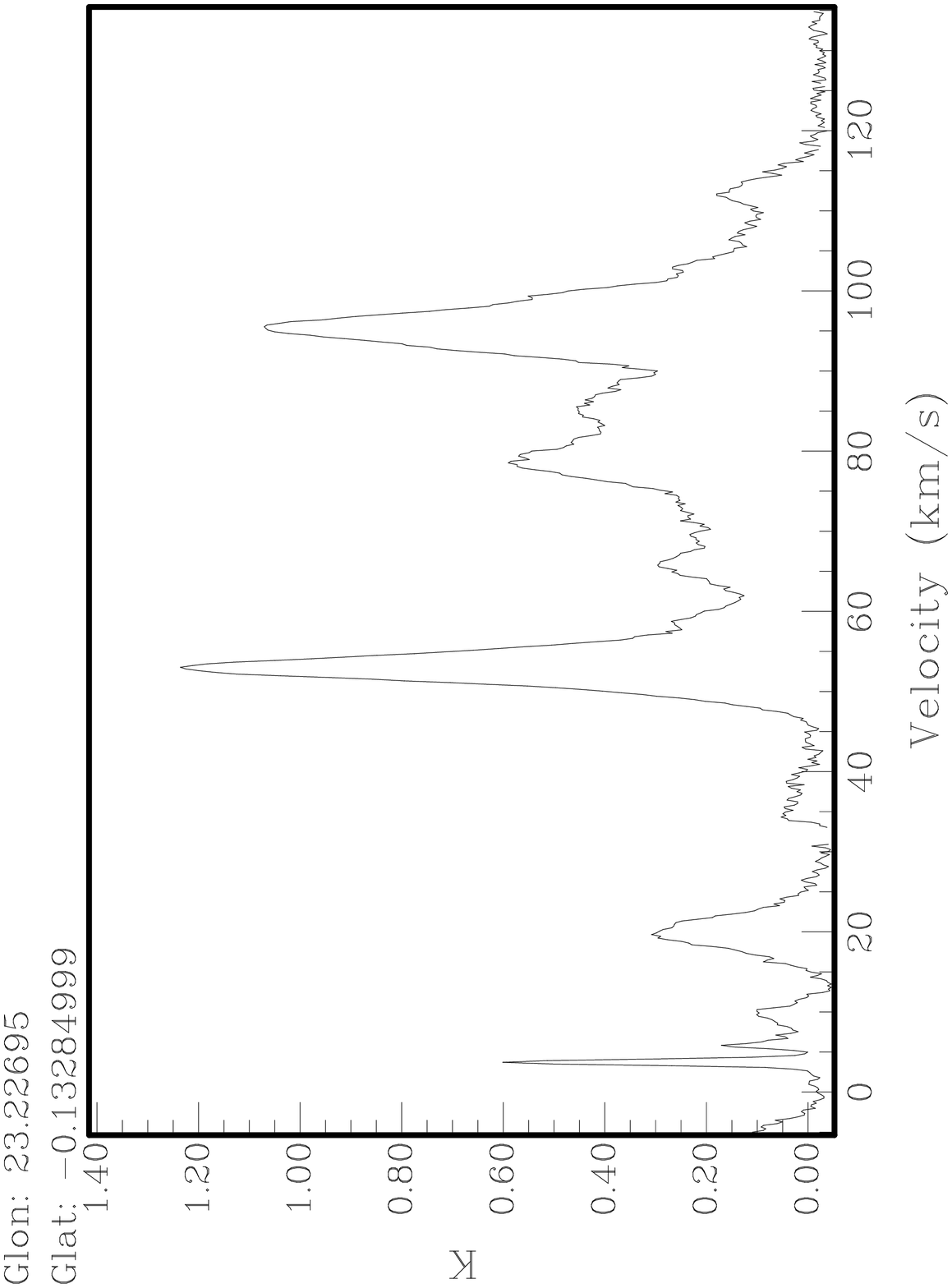}}
\put(260,320){\includegraphics{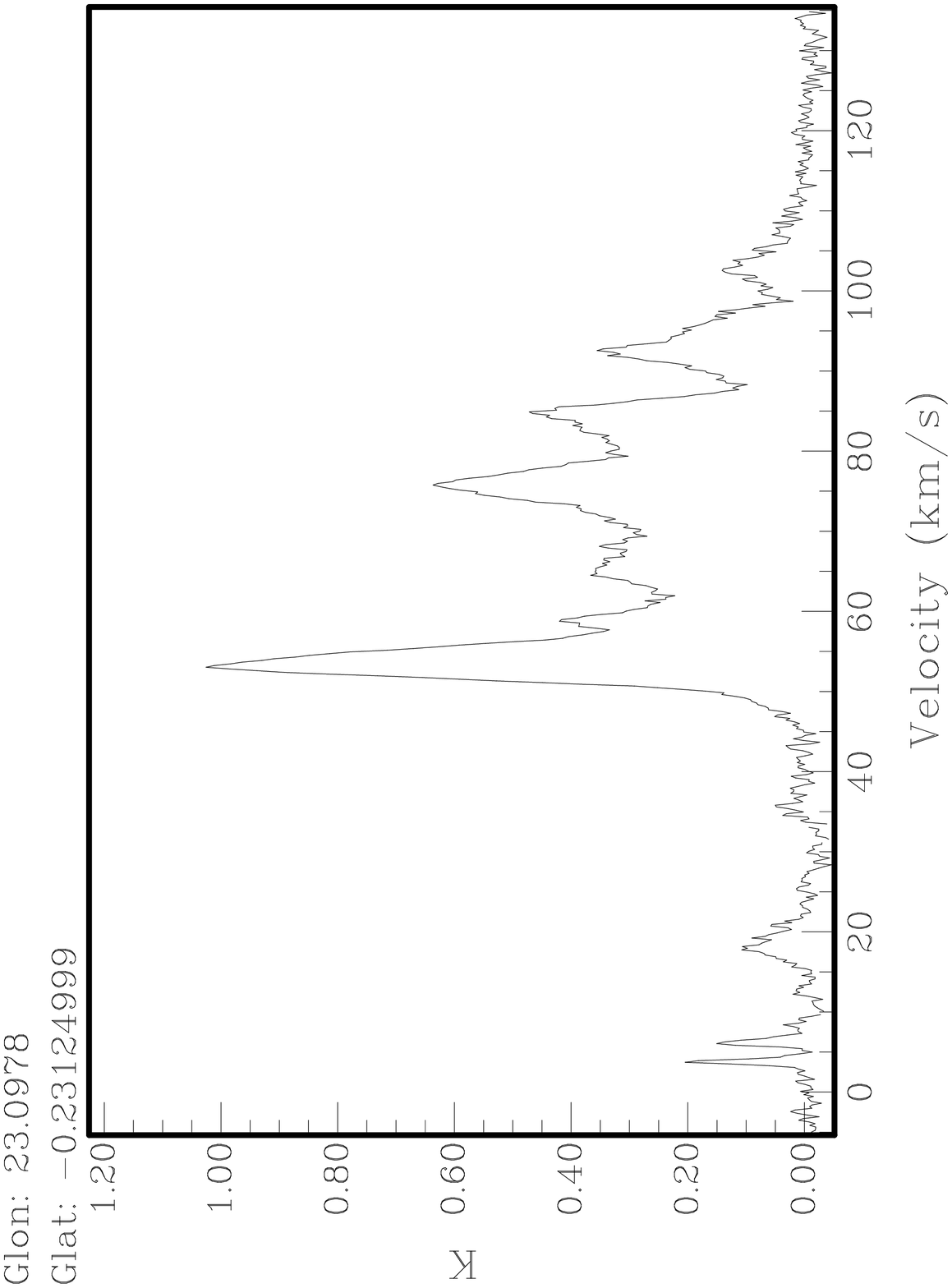}}
\put(0,155){\includegraphics{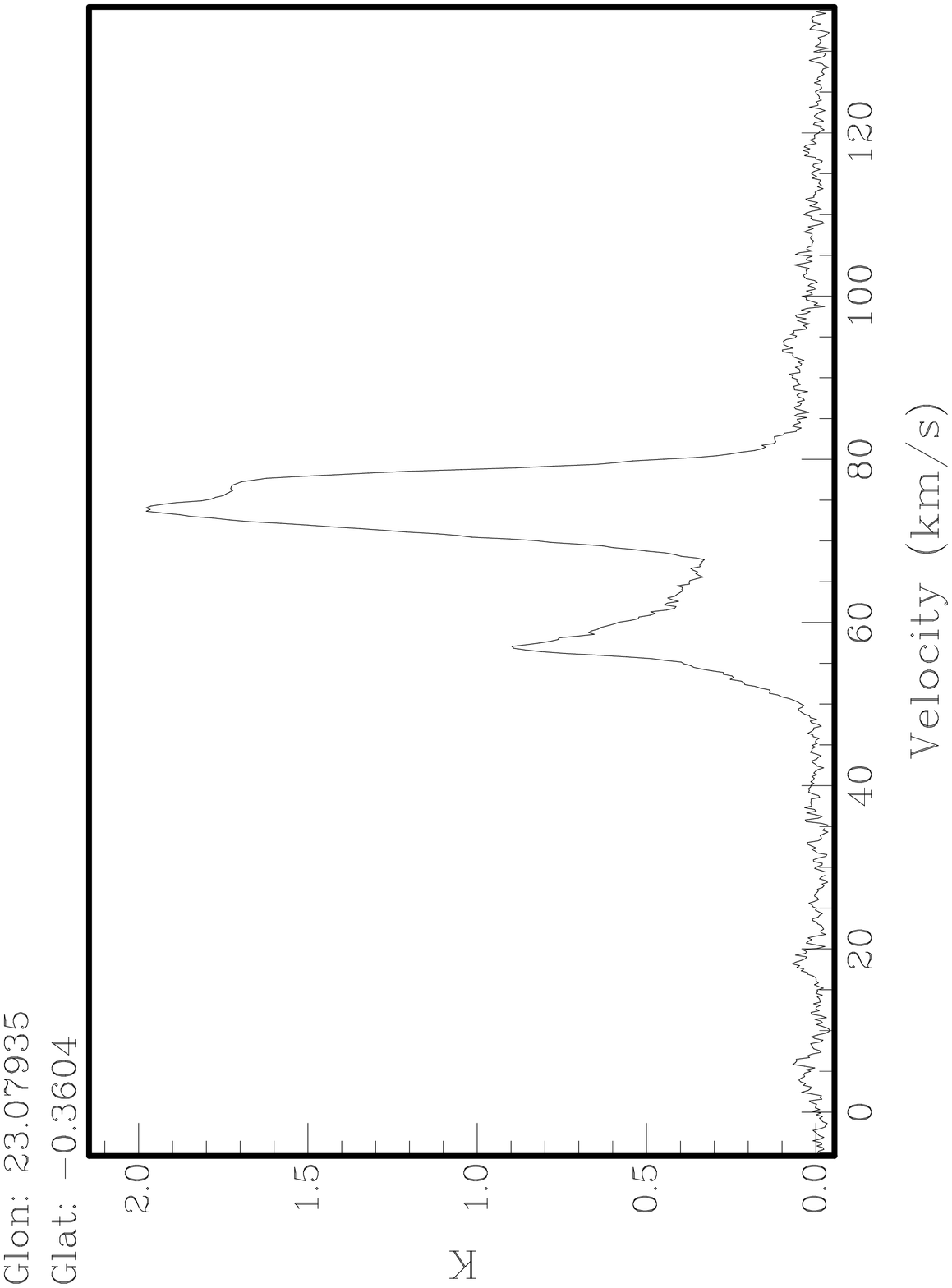}}
\put(260,155){\includegraphics{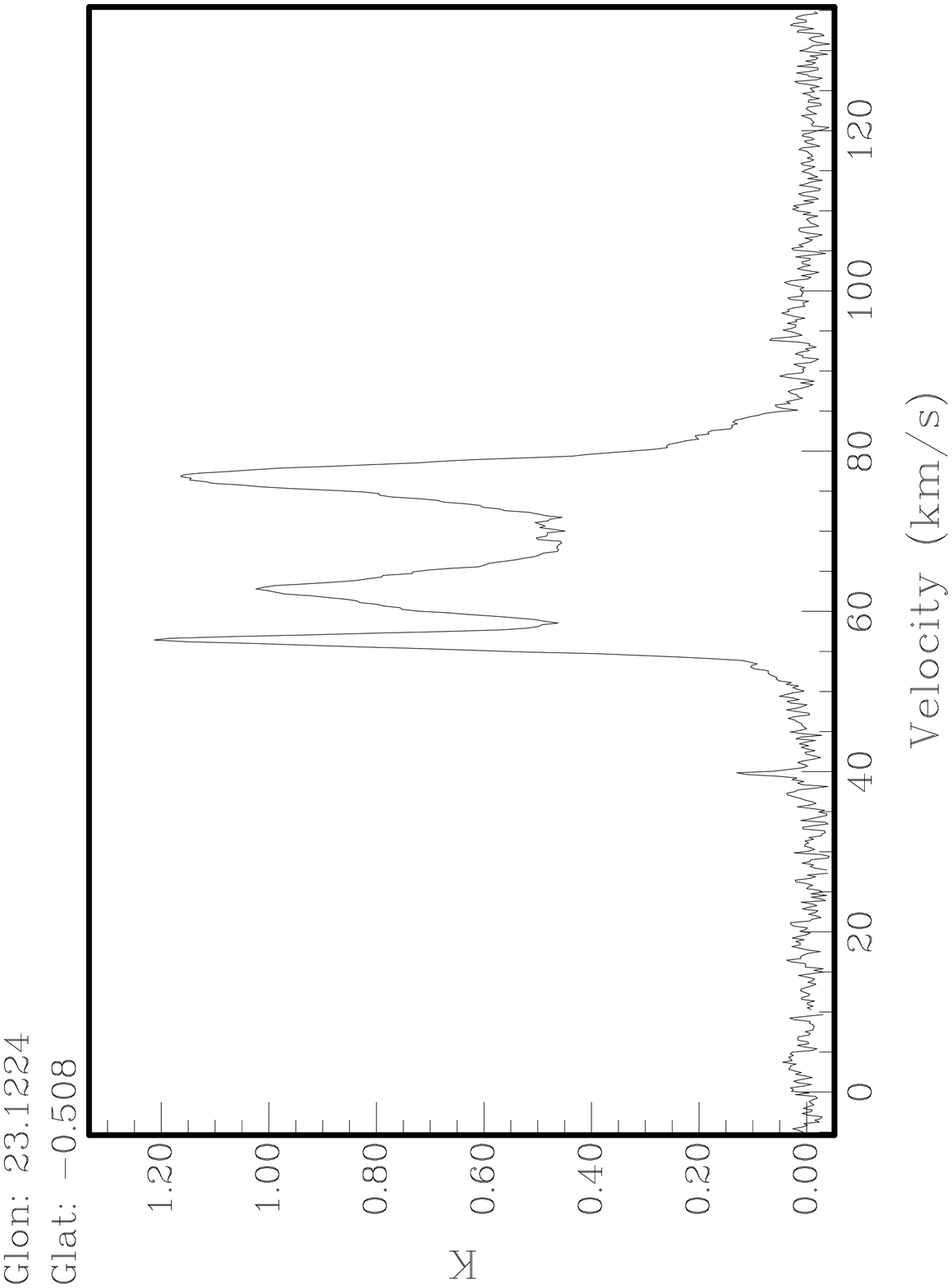}}
\end{picture}
\caption{The 8 panels show the CO spectra extracted from the regions C1, C2 (top row), 2, 3 (second row), 4, 5 (third row), 6 and 7 (last row)}
\end{figure}

\end{document}